\documentclass[letter,preprint]{jpsj2} 
\def\Vec#1{\mbox{\boldmath $#1$}}
\usepackage{graphicx}
\usepackage{dcolumn}
\usepackage{bm}

\title{Non-Gaussian Velocity Distribution Function in a Vibrating Granular Bed}

\author{Atsushi \textsc{Kawarada}\thanks{E-mail address: ak@yuragi.jinkan.kyoto-u.ac.jp} and Hisao \textsc{Hayakawa}\thanks{E-mail address: hisao@yuragi.jinkan.kyoto-u.ac.jp}}

\inst{Department of Physics, Yoshida-South Campus, Kyoto University,
Kyoto 606-8501}
\recdate{\today}

\abst{The simulation of granular particles in a quasi-two-dimensional
 container under vertical vibration as an experimental accessible
 model  for granular gases  is performed. 
The velocity distribution function obeys an exponential-like function 
during the
 vibration and deviates from the exponential function in free-cooling states.
It is confirmed that this exponential-like distribution function is produced
 by Coulomb friction  force. A Langevin equation with Coulomb
 friction is proposed to describe the motion of such a system.}

\kword{granular gas, velocity distribution function, non-Gaussian, flatness }

\begin{document}
\maketitle

Granular physics has been a  challenging field in statistical
physics since the rediscovery of significant nature in granular
materials in the late '80s or early '90s.\cite{jaeger}
An assembly of grains has strong fluctuations in the configuration 
and the motion such that  mean-field theories cannot be used in most 
situations. A typical example of the strong fluctuation appears in the
force distribution for a static granular assembly piled by 
gravity.\cite{danta,mueth,mueth2}
 The
force propagates along force chains and the distribution function of
the magnitude of the force on the bottom of a container does not obey 
a Gaussian function but an exponential function.\cite{danta,mueth,mueth2}   

Such a strong fluctuation originaling from non-Gaussian properties
should be relevant even in the dynamics of
granular assemblies.
However, there have not been many systematic
research studies that focus on the statistical distribution functions in  
 the  steady state,
because real and numerical experiments report no unified results, i.e., 
velocity distribution
functions (VDFs)  obey 
Gaussian-like functions with exponential tails\cite{gas1,gas12,gas13,gas2,gas3,gas4,warr,warr2},from the Gaussian to the exponential depending on 
density\cite{murayama,olafsen}, stretched exponential\cite{losert,menon,moon} 
and even power-law functions\cite{kudrolli,taguchi,taguchi2,taguchi3}.

As in standard statistical mechanics, an assembly of grains in a gas phase is
an idealistic system for studing 
statistical weight.
Approximate granular gases can be obtained by rapid granular flows on 
inclined slopes\cite{namiko,namiko2}, gas-solid mixtures\cite{tanaka,tanaka2} and
grains under the external
vibration.\cite{warr,warr2,murayama,olafsen,losert,menon,moon,kudrolli} However, these
systems cannot be regarded as idealistic granular gases, because 
(i) the boundary and gravity effects are strong in the rapid
granular flows, (ii) the hydrodynamic interaction between particles in
 gas-solid mixtures are complicated, and (iii) a dense cluster appears in the vibrating
experiment.  Therefore, it is difficult to achieve
free-cooling gases in experiments.\cite{losert}

In this letter, to remove such problems, 
at first, we propose an experimental accessible situation to produce
granular gases. Second, 
we demonstrate that VDFs in both dense and 
dilute granular gases under vertical  vibration 
can obey exponential-like functions when Coulomb friction  is
important through our simulation based on the distinct element method (DEM). 
This exponential VDF disappears immediately after
 vibration has stopped, i.e., the
grains are in a free-cooling process.
Third, we introduce a phenomenological Langevin equation eq. (\ref{langevin}) to describe the 
motion of particles and explain the mechanism that causes the exponential VDF 
to appeas.
The detailed analysis  of experimental accessible granular
gases will be reported  elsewhere.

We use a three-dimensional DEM for monodispersed 
spherical particles.\cite{cundall} Instead of using the Hertzian contact
force, we adopt the linear spring model to represent the repulsion of
contacted spheres. We also include the rotation of spheres and Coulomb slip for tangential contact. The equations of motion of
the sphere  with the diameter $d$, the mass $m$ and the momentum of
inertia $I=md^2/10$ at contact are given by
\begin{eqnarray}\label{eqofmo}
m \ddot{\bf x}_i&=&\sum_{<ij>}\{{F_n}^{ij}{\bf n}_{ij}+\tilde{F_t}^{ij}{\bf t}_{ij}\}
-m {\bf g} ,\nonumber\\
I \dot {\Vec{\omega}}&=& \frac{d}{2}\sum_{<ij>}\tilde{F_t}^{ij}
\{{\bf n}_{ij}\times {\bf t}_{ij}\} ,
\end{eqnarray}
where ${\bf n}_{ij}$, ${\bf t}_{ij}$, $\sum_{<ij>}$ and ${\bf g}$ 
represent the normal and the tangential unit vectors at the contact point 
of the $j$ and $i$ particles,
the summation over $j$ particles which
are in contact with $i$ particle, and the gravitational acceleration, 
respectively. The contact forces ${F_n}^{ij}$ and $\tilde{F_t}^{{ij}}$ 
are respectively given by
\begin{eqnarray}\label{force}
 {F_{n}}^{ij}&=&
-k_n(d-|{\bf x}_i-{\bf x}_j|) -m\eta_n {\bf n}_{ij}\cdot
(\dot{\bf x}_i-\dot{\bf x}_j), \nonumber \\
\tilde{{F}_t}^{ij} & = &
\left\{
      \begin{array}{rl}
        {F_t}^{ij} & ({\rm if} \quad |{F_t}^{ij}| < \mu |{{F}_n}^{ij}|) \\
        \mu {{F}_n}^{ij} & ({\rm otherwise}) 
      \end{array}
\right. ,\nonumber \\
\end{eqnarray}
where 
\begin{equation}
{F_{t}}^{ij}=-k_t(d-|{\bf x}_i-{\bf x}_j|)-m\eta_t {\bf t}_{ij}
\cdot(\dot{\bf x}_i-\dot{\bf x}_j+\frac{d}{2}(\Vec{\omega}_i+\Vec{\omega}_j)).
\end{equation}
Here, the friction coefficient $\mu$ describing Coulomb slip is assumed
to be $\mu=0.8$. In addition, 
we introduce four parameters $k_n, k_t, \eta_n$ and $\eta_t$ which are
assumed to be $k_n=5.0\times 10^4mg/d$, $k_t=2.5\times 10^4mg/d$,
$\eta_n=4.47\times 10 \sqrt{g/d}$ and $\eta_t=\eta_n/2$ in our
simulation. This setup corresponds to the normal restitution coefficient 
$e=0.638$. We adopt the second-order
 Adams-Bashforth method for the time-integration
with the time interval $\Delta t=1.0 \times 10^{-4}\sqrt{d/g}$.

We focus on the following situation. Particles are confined in a
quasi-two-dimensional container in which the height is 1.8$d$ and
the horizontal plane is a square (Fig. 1). In this system, particles
cannot from a multilayer configuration in the vertical direction, but 
the particles have vertical velocities in their motion. 
The mobile particles are
randomly scattered by fixed particles on the top of the container if the 
container is vertically vibrated. 
The area fractions of fixed scatters are 0.21, 0.23 and 0.25 for 1,000, 3,000 and 10,000 mobile particles, respectively.
The vibration is driven by a sinusoidal 
force whose  amplitude and frequency are given by $A=1.2d$ and 
$f=0.5\sqrt{g/d}$. Thus, the acceleration
amplitude becomes $\Gamma=A(2\pi f)^2/g=11.5$.
If the vibration is stopped, the
particles move with the rotation on the bottom plate, 
then the system can simulate a two-dimensional free-cooling granular
gas. We emphasize that the rotational particles on a flat plane can be 
regarded as an approximate 
assembly of gas particles in a two-dimensional system,
because the energy loss in the rolling friction is very small in many cases.

\begin{figure}
 \begin{center}
  \includegraphics[width=78mm]{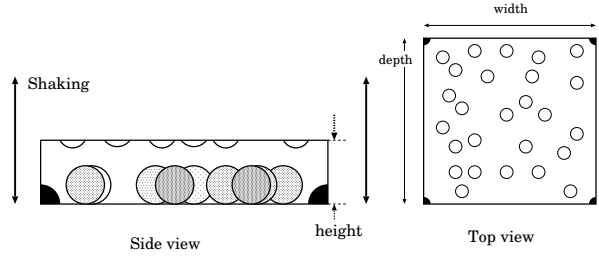}
 \end{center}
\caption{
Schematic 
side view (left) and top view (right) of our setup. The random
 scatters are fixed spherical particles on the top board. The diameter
 of fixed particles on the top plate (white ones) is uniformly random
 between $0.6d$ and $0.8d$,  and their centers are located $0.15d$ above the
 top board.  To avoid the stacking of particles in the corners, we introduce
 four fixed particles (black ones in the figure) at the corners.
}
\end{figure}

The simulation starts from a stationary state of particles on the
bottom plate. The particles gain the kinetic energy from the external
oscillation and the system reaches a `steady state' in the balance between
collisional dissipations and the gain of the energy from the external force. 
Typically, the system reaches the statistical 
`steady state' after 25 oscillation cycles 
in which hydrodynamic quantities are independent of how
many cycles proceed from the initial condition but depend only on the
phase of oscillation. 

In the `steady state', we confirm that the density is uniform and 
there is no long-range
correlation in contrast with granular gases in free-cooling
states.\cite{tpj} Thus, the system does not have any systematic flows nor any 
definite clusters. We also check that the effects of side boundaries can be
neglected. This is because particles always hit the top wall or the bottom wall
during the vibration.
After we stop the vertical
oscillation, the correlation
grows with time as the free-cooling process proceeds.   

\begin{figure}
 \begin{center}
  \includegraphics[width=78mm]{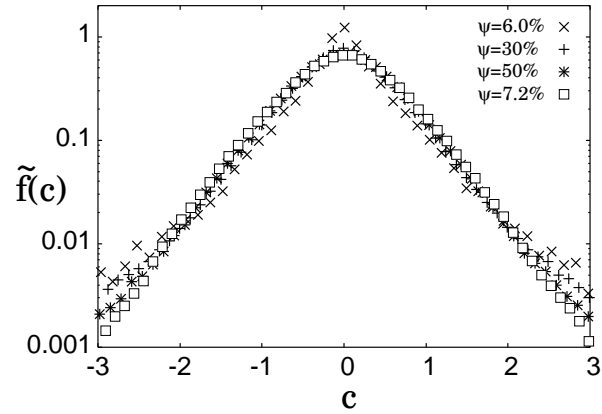}
 \end{center}
\caption{
Scaled VDF $\tilde f(c)$ in `steady states' under the 
 vertical vibration. Here, $\Psi$ represents the area fraction. For 
$\Psi=7.2\%$ we use 10,000 particles, and use 3,000 particles for other cases.
}
\end{figure}

\begin{table}
  \caption{Flatnesses(FN) under several conditions. See text for details.}
  \begin{center}

    \begin{tabular}[]{lcrrc}
      \hline
      & Number of & \multicolumn{2}{c}{Fractions(\%)}& FN \\
\cline{3-4}
      & particles &      Area & Volume &  \\
\hline
\hline
steady state & $1.0\times 10^3$  &  $6.5$  & $2.4$  & $8.00$  \\
          & $$      &  $10$  & $3.7$  & $7.63$  \\
          & $$      &  $20$  & $7.4$  & $6.90$  \\
          & $$      &  $30$  & $11$  & $6.26$  \\
          & $$      &  $40$  & $15$  & $5.67$  \\
          & $$      &  $50$  & $19$  & $5.36$  \\
\cline{2-5}
          & $3.0\times 10^3$  & $6.0$  & $2.2$  & $9.57$  \\
          & $$       & $10$  & $3.7$  & $8.13$  \\
          & $$       & $20$  & $7.4$  & $7.03$  \\
          & $$       & $30$  & $11$  & $6.34$  \\
          & $$       & $40$  & $14$  & $5.64$  \\
          & $$       & $50$  & $18$  & $5.27$  \\
\cline{2-5}
          &$1.0\times 10^4$  &  $7.2$  & $2.6$ &  $6.85$
 \\
\hline 
free-cooling &$1.0\times 10^4$ & $7.2$  & $2.6$&  $4.20$  \\
\hline 
no tangential   & $1.0\times10^3$  & $6.5$  & $2.4$  & $3.26$  \\
                 & $3.0\times 10^3$ & $6.0$  & $2.2$  & $3.43 $  \\
\hline 
no friction      & $1.0\times10^3$  &  $6.5$     & $2.4$      & $3.27$  \\
                 & $3.0\times 10^3$ &  $6.0$     & $2.2$      & $3.43$  \\
\hline 
undulation      & $1.3\times 10^3$ &  &         & $4.10$  \\
\hline

    \end{tabular}
  \end{center}
  \label{table:flatness}
\end{table}
The most significant quantity for charactering 
 this gas system is VDF. In the `steady state', the vertical
component of VDF 
has double peaks  where each peak corresponds to a lifting process  or
a falling process, while the horizontal VDF has a single peak.
For later discussion, we are only interested in the horizontal VDFs for 
analysis. We plot the scaled horizontal VDF $\tilde f(c)$ in Fig. 2 using
\begin{equation}
f({\bf v},t)=nv_0(t)^{-2}\tilde f({\bf v}/v_0)
\end{equation}
with the density $n$, the average speed $v_0=\sqrt{2T/m}$ with the
granular temperature $T$,
 and $\int d{\bf c}\tilde f(c)=\int d{\bf
c}c^2\tilde f(c)=1$, where $\tilde f(c)$ is averaged over the cycles.
As in Fig. 2,
the scaled VDF can be approximated by an exponential function. In fact, 
the flatness defined by $<c^4>/<c^2>^2=<c^4>$ with 
$<c^n>=\int d{\bf c}c^n\tilde f(c)$ is not far from 6.
It should be noted that the flatness with the 
Gaussian VDF is 3 and that with the exponential VDF is 6.
In our simulation, the flatnesses 
are summarized in Table I for
1,000- and 3,000-particle simulations as functions of area fractions projected into the horizontal plane. 
In addition, the dependence of the flatness on the density is
relatively weaker in our situation than that by Murayama and Sano\cite{murayama}. 
A large system with 10,000 particles with an area
fraction of 7.2\% has a smaller flatness of 6.85. 
If the external oscillation stops, the flatness decreases quickly and is 
 saturated at 4.20 for 10,000 particles.
As can be seen in Fig. 3, VDF in the cooling process is nearly Gaussian 
for low-energy particles but has an exponential tail for high-energy
 particles as in the usual gas systems.\cite{gas1,gas12,gas13}

Let us consider the origin of the exponential-like VDF.  It is easy to verify 
that the exponential VDF cannot be reproduced without the existence of
Coulomb friction in eq. (\ref{force}). Namely, if we eliminate the tangential
component of the contact force or the slip rule for 
$|{F_t}^{ij}| < \mu |{{F}_n}^{ij}|$, VDF becomes a Gaussian-like
distribution (Fig. 3). In fact, if all the components of tangential contact
force are omitted, the flatness becomes 3.43 for 3,000 particles with the 
area fraction of 6.0\%. While if we only neglect the effect of Coulomb 
friction in eq. (\ref{force}), the flatness becomes 3.43 in the same situation(Table \ref{table:flatness}).
In our system, particles experience an strong shear force when they hit 
the 
fixed scatters on the top wall, 
and their direction of motion changes drastically. Thus, the tangential slips between particles and
the fixed scatters are the dominant
dissipative processes in the `steady state'.
Since we specify the origin of exponential-like VDF, we can understand
the weak dependence of VDF on the density. Namely, particles 
directly hit the fixed scatters for dilute case, while dense particles
collide with each other
and  rotate without
slips besides the collisions with the walls.

\begin{figure}
 \begin{center}
  \includegraphics[width=78mm]{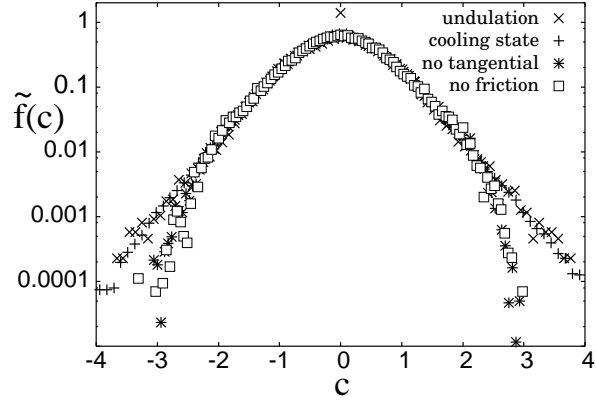}
 \end{center}
\caption{
Scaled VDF for several situations. In the inset,
`no tangential' and `no friction' correspond to the results of
 simulations without the tangential force and the friction force,
 respectively, where the systems are in the `steady state' and include
 3,000 particles with the area
 fraction 6.0\%. Here, `cooling state' and 
`undulation' indicate the results of the simulation of the cooling process
 of 10,000 particles and the undulation, respectively.
}
\end{figure}

 Therefore, the essence to produce the large flatness in VDF is
apparently Coulomb friction. 
Thus, we propose the following Langevin equation to describe
the horizontal motion of particles.
\begin{equation}\label{langevin}   
\frac{d{\bf u}}{dt}=-\gamma \frac{{\bf u}}{u}-\nabla\Phi+\Vec{\eta}
\end{equation}
Here, ${\bf u}$, $u=|{\bf u}|$ and $\Phi$ are respectively 
 the velocity, the speed and the potential exerted among particles. 
The friction coefficient $\gamma$ may be proportional to $\mu g$.
The $\alpha$ component of the random force
$\Vec{\eta}$ satisfies the fluctuation-dissipation relation.
\begin{equation}\label{FDT}
<\eta_{\alpha}(t)>=0,\quad 
<\eta_{\alpha}(t)\eta_{\beta}(t')>=2D\delta_{\alpha\beta}\delta(t-t')
\end{equation}
Here, $D=\gamma\sqrt{T/3m}$ is
the diffusion coefficient in the velocity space.
The Langevin equation eq. (\ref{langevin}) with eq. (\ref{FDT}) can be converted into 
the Fokker-Planck or Kramers equation for the probability 
distribution function $P({\bf x},{\bf u},t)$.
\begin{eqnarray}\label{FP}
\frac{\partial P({\bf x},{\bf u},t)}{\partial t}
&=&\{
-\nabla \cdot {\bf u}+
\frac{1}{m}\frac{\partial}{\partial {\bf u}}\cdot\nabla \Phi+ \nonumber \\
& &
\gamma\frac{\partial}{\partial {\bf u}}\cdot \frac{{\bf u}}{u}+
D\frac{\partial^2}{\partial u^2}
\}
P({\bf x},{\bf u},t)
\end{eqnarray}
For spatial homogeneous `steady states', the equation
for $P\to P_{st}$ 
is reduced to $ ({\bf u}/u)P_{st}+ \sqrt{T/3m}(d/d{\bf u})P_{st}=0$ for
${\bf u}\ne 0$. Its solution can be obtained easily
\begin{equation} 
P_{st}({\bf u})=2\displaystyle\sqrt{\frac{m}{3T}}\exp[-\displaystyle\sqrt{\frac{3m}{T}}u].
\end{equation}
Thus, the Langevin equation with Colomb friction law obeys the
exponential VDF.

To check the validity of the new Langevin equation for the motion of 
particles, 
we evaluate both the diffusion coefficient $D$ and the friction
coefficient $\gamma$. At first, we have confirmed that the motion of
particles in the horizontal plane is diffusive. 
Then, the diffusion coefficient is evaluated from
the relation $<({\bf r}(t)-{\bf r}(t_0))^2>=4D(t-t_0)$, where we choose 
$t_0$ as  25 cycles of the oscillation and simulate the motion of particles
until 75 cycles. For each parameter of
the oscillation we determine $T$ from the second
moment of VDF. Thus, we obtain the relation between $D$ and $T$ for simulations of 1000 and 3000 particles as in Fig. 4. 
The best fitted relation is $\tilde D=0.0735 \tilde
T^{0.452}-0.0078$, where $\tilde D=D/(d^{1/2}g^{3/2})$ and $\tilde
T=T/(mgd)$, but this can be replaced by $\tilde D=0.0744 \sqrt{\tilde
T}-0.0054$. This result seems to be independent of the system size. 
We also note that the collective vertical motion suppresses
the diffusion for larger $T$, and no diffusion takes place because of the insufficient
kinetic energy for smaller $T$.
From the relation between $\tilde D$ and $\tilde T$, we evaluate $\gamma$ 
as $\gamma=0.129g$. Note that $\gamma$ is independent of $T$, because 
the dominant dissipative process involves collisions between
particles and the horizontal walls caused by the collective motion
of particles in the vertical direction. Thus, we believe that the
Langevin equation eq. (\ref{langevin}) can be used to describe the motion of 
particles.

\begin{figure}
 \begin{center}
  \includegraphics[width=78mm]{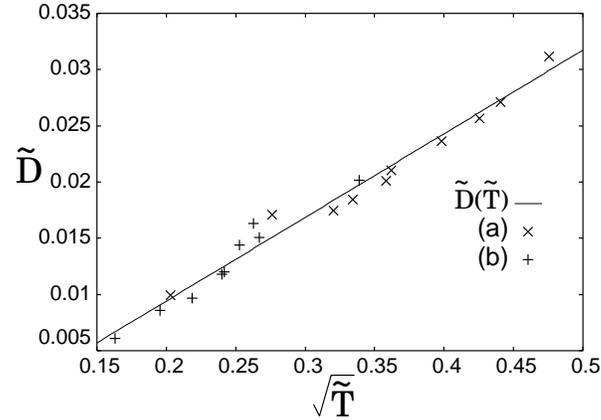}
 \end{center}
\caption{
Relationship between $\tilde D$ and $\tilde T$. The solid line
 represents $\tilde D=0.0744\sqrt{\tilde T}-0.0054$. (a) 
and (b) represent the data for 3,000 particles and 
1,000 particles,respectively.
}
\end{figure}

Figure 3 also contains the VDF obtained from the
simulation of granular particles confined in a dense cluster 
which exhibits undulations.
In the simulation, we
use $k_n=2.0\times 10^4mg/d$, $k_t=k_n/2$, $\eta_n=\eta_t=3.8\times
10\sqrt{g/d}$ and $\mu=0.4$ with polydispersed particles whose diameters 
are uniformly distributed between 0.8$d$ and $d$. The number of particles is
1,300 and confined in a thin box whose depth is 4.5$d$ and width is
60$d$. The initial accumulated height of particles  is approximately 6 layers. 
 In this situation, 
the relative motion of particles is suppressed and particles
are confined in a lattice-like cage. In this case, VDF is deviated
from the exponential function where the flatness is approximately 4.10(Table \ref{table:flatness}), but the high-energy particles 
apparently obey an exponential law as in the
free-cooling case(Fig. 3). It is easy to
imagine that particles with a high energy can slip on the cage and
particles with a low energy oscillate within cages. 
This picture may be valid for the gas system in the free-cooling process.
Thus, VDF can be
approximated by the combination of the 
Gaussian part for low-energy particles and 
the exponential part for
high-energy particles. 
The VDFs for the undulation and the free-cooling process 
are similar to those in experiments under 
vertical vibration in which a dense cluster exists and each VDF can be
approximated by a single stretched exponential
function\cite{warr,warr2,losert,menon,moon,kudrolli,olafsen}. We consider,
however, that the stretched exponential VDF can be understood 
by the combination of the Gaussian part and the exponential 
part as stated here.

We also recall that VDF 
obeys a Gaussian-like function for dense flows on an inclined 
slope.\cite{mitarai} 
This
is because particles can move without slips because the lattice-like 
cage is very weak.  We also compare our results with those of Murayama and
Sano\cite{murayama}. Because of the nonexistence of random scatters in their
simulation, the slip processes can occur as results of 
collisions among particles. Thus it is reasonable to get the transition from
the  Gaussian to the exponential as the density increases. On the other
hand, in our case, the slips take place mainly in collisions between
particles and fixed random scatters. 
Therefore the tendency to obey the exponential is 
emphasized for dilute gases.

This mechanism to obtain the exponential distribution function may be
used for  other situations, because Coulomb friction plays roles
in sliding frictions \cite{persson}. To look for possibilities of the universal feature
of our scenario will be our future task. 

In conclusion, we propose an experimental accessible system for granular
gases. The VDF in a `steady state' obeys an exponential-like function but
changes to Gaussian-like distribution function when free-cooling starts.
This exponential VDF is caused by Coulomb friction force. Thus, we
propose the Langevin equation with Coulomb friction to reproduce the
results of our simulation.

\begin{acknowledgments}

HH thanks Y. Murayama and N. Mitaraithe for fruitful discussion. 
This work is partially supported by the Grant-in-Aid for Scientific 
Research (Grant No. 15540393) of the Ministry of Education, Culture, Sports, 
Science and Technology, Japan.

\end{acknowledgments}

\end{document}